# Manually Tunable Ventilated Metamaterial Absorbers

Xiao Xiang[1], Hongxing Tian[1], Yingzhou Huang[1,a)], Xiaoxiao Wu[1,2,a)], Weijia Wen[2,3]

[1]*State Key Laboratory of Coal Mine Disaster Dynamics and Control, Chongqing Key Laboratory of Soft Condensed Matter Physics and Smart Materials, Chongqing University, Chongqing, 400044, China*

[2]*Department of Physics, The Hong Kong University of Science and Technology, Clear Water Bay, Kowloon, Hong Kong, China*

[3]*Materials Genome Institute, Shanghai University, Shanghai 200444, China*

[a)]Authors to whom correspondence should be addressed: yzhuang@cqu.edu.cn (Y. Huang) and xwuan@connect.ust.hk (X. Wu).

## Abstract

For most acoustic metamaterials, once they have been fabricated, their operating frequencies and functions cannot be adjusted, which is an intrinsic barrier for development of realistic applications. The study to overcome this limit has become a significant issue in acoustic metamaterial engineering. Although with the advance of metamaterials in the past two decades, a series of methods such as electric or magnetic control have been proposed, most of them can only work in the condition of no fluid passage. Some metamaterials with large transmission losses have been proposed, but the sounds are essentially reflected rather than absorbed. Here, to overcome this intrinsic difficulty, we propose a ventilated sound absorber that can be manually tuned in a large range after being manufactured. During the tuning which is achieved through an intricately designed slider, high-performance absorption and ventilation are both ensured. The tunable ventilated sound absorber is demonstrated experimentally and the effective model of coupled lossy oscillators can be employed to understand its mechanism. The manually tunable ventilated metamaterial has the potential application values in various complicated pipe systems that require frequency adjustment and it also establishes the foundation for future development of active tunable ventilated acoustic metamaterials.

In the past two decades, acoustic metamaterials have developed rapidly in various fields such as acoustic cloaking[1,2], subwavelength imaging[3-5], topological acoustics[6-9], and sound insulation and absorption[10,11], providing unprecedented ways to control sound waves, which have fundamental theoretical and application values[12-15]. For example, the rapid development of sound-absorbing metamaterials has solved the critical issue of the low dissipation of traditional sound-absorbing structures under low-frequency sound waves (< 500 Hz) which resulting in inefficient absorption of sound energy[13,14,16-19]. However, these acoustic metamaterials cannot work with free fluid such as air. Some ventilated acoustic metamaterials have been proposed previously but cannot address difficulties of absorption at low frequencies[20-32]. Furthermore, the majority of acoustic metamaterials that have been proposed and demonstrated in the past two decades are passive with fixed geometries. Once they have been fabricated, their operating frequencies and functions cannot be tuned. They cannot adapt to complex environment where noise frequencies are usually unknown. To solve this issue, varieties of tunable acoustic metamaterials have been proposed and investigated in recent years. Some tunable mechanisms have been presented, mainly employing mechanical deformations, piezoelectric, and/or magnetic effects[33-36]. However, they all require rigid backings to terminate sound transmissions, which also forbid transmissions of fluids, such as air and water. Further, their complex and intricate mechanical, electronic, and/or magnetic structures are not convenient and robust for everyday applications, and a simple adjustment method are much preferred.

In this Letter, a manually tunable ventilated metamaterial absorber (MTVMA) are proposed and experimentally demonstrated. As mentioned, previous tunable acoustic metamaterial absorbers only work in the condition of no transmissions and generally employ the electromechanical or magnetomechanical effect which requires external circuits and structures to be effectively adjusted. In contrast, the MTVMA can work in an open and ventilated environment, and can be simply manually adjusted to suit for different working frequencies. Therefore, it can simultaneously achieve high-performance acoustic absorption (>85% at resonance) and ventilation (>50% wind velocity ratio) within a large range of tunable working frequencies. In brief, the MTVMA is comprised of carefully designed weakly coupled Helmholtz resonators[21,37-40], and its absorption is demonstrated through numerical calculations and experimental measurements. It provides a direct route for simultaneously achieving tunable working frequency through adjustable geometry and high-efficiency absorption with ventilation at low frequencies. The practical applications can be found in various acoustic engineering scenarios where open environment must be accommodated while in dealing with varieties of noises.

The MTVMA unit is comprised of two identical Helmholtz resonators with the spatial inversion symmetry, and the units are assembled as schematically illustrated in Fig. 1(a). A large hollow in the upper part of the frame permit background fluids (here, the air) freely passing through the structure. In this work, it is assumed that the all MTVMAs are placed in air. The absorber demonstrated here comprises four MTVMA units arranged in a rectangular lattice, and the lattice constants of MTVMAs are $l$ and $l/4$ along the $x$ and $y$ directions. The working frequency of MTVMAs can be continuously tuned by moving the slider, as is illustrated in Fig. 1(b), where the covers are removed to demonstrate inside details of the MTVMA unit (note that the structure is rotated by 90° for better visualization), and the slider parts are denoted with blue color. The detailed sectional views on middle $yz$-plane of a single MTVMA unit are depicted in Fig. 1(c). The MTVMAs are fabricated by using a three dimensional (3D) printer with an accuracy of 0.1 mm. The printing material is made of photosensitive resin composite, which has a tensile modulus 2.46 GPa and density 1.10 g/cm$^3$ after being cured. For characterizing sound absorption efficiency of the samples, we perform measurements in an impedance tube with a square cross section using the four-microphone method as is shown in Fig. 1(d)

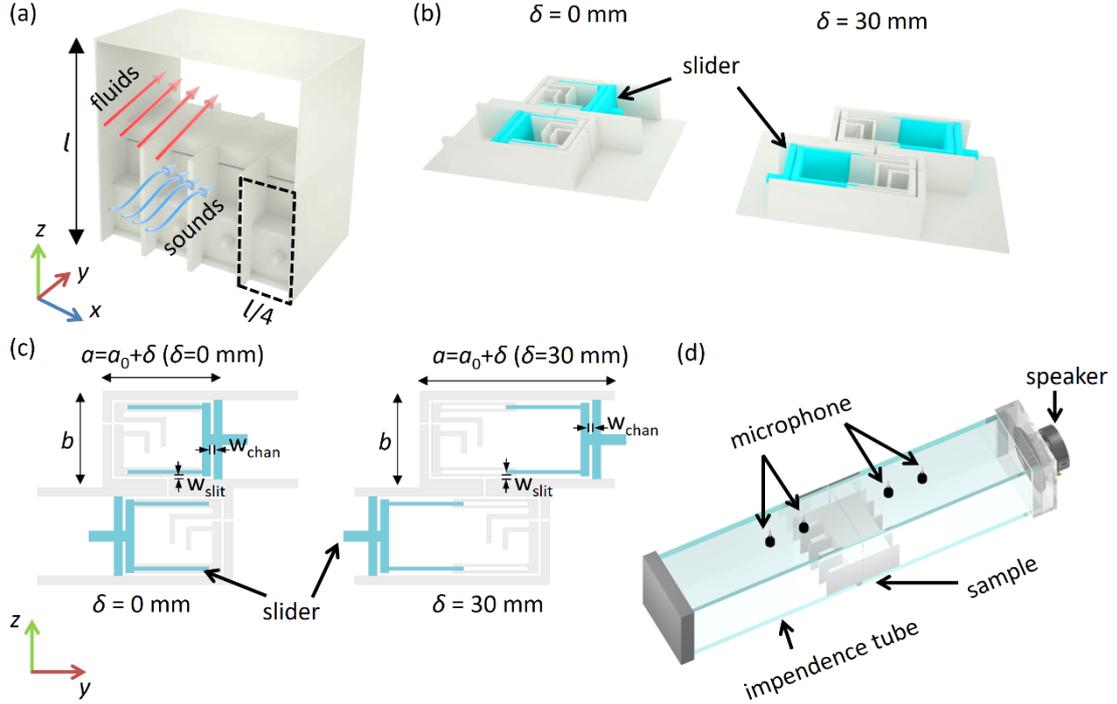

Fig. 1. (a) Supercell in a rectangular lattice consisted of four MTVMA units. The lattice constants along $x$ and $z$ directions are $l/4$ and $l$, respectively. (b) Close view of a single MTVMA unit. It transits from the minimum position ($\delta$ = 0 mm) to the maximum position ($\delta$ = 30 mm) by adjusting the slider (see Movie S1 in supplementary material). To demonstrate inside details, the structure is rotated and the cover is removed. The slider parts are denoted with blue color. (c) Sectional schematic of the MTVMA on the $yz$-plane. (d) Experimental setup for acoustic measurement. The square cross-section of the impendence tube is 147×147 mm$^2$ and a fabricated sample is placed in the impedance tube, and we adopt the four-microphone method for all the acoustic measurements.

In order to adjust the slider to reconstruct acoustic metamaterials, the relationship between tunable frequency and MTVMA geometry parameter is investigated. All simulations are performed with a commercial finite-element method (FEM) software COMSOL Multiphysics (see supplementary material for setup details). Among them, $d_{chan}$, $d_{slit}$, $a$, $b$ are key parameters that have major impact on the absorption frequency and absorption rate. Among these parameters, we chose the parameter $a$ which has a larger adjustable range and a significant influence on the sound absorption frequency as the adjustment parameter. Therefore, the parameter $a$ is considered while maintaining the other parameters ($b$ = 45 mm, $d_{chan}$ = 1.4mm, $d_{slit}$ =1.4mm, $d_{wall}$ = 2mm). The absorption spectrum of parameter $a$ is shown in Fig. 2(a). The bright strip in the color map highlights the shift of the resonance of the MTVMA unit. As the length $a$ is increased, the resonance shifts towards low frequency, from 500Hz to 300Hz, and the absorption peak retains. The energy of common noises in daily life, such as those from many household appliances and ducts, starts to become significant when the frequency is higher than 300 Hz[41-43]. Therefore, it is important to realize an acoustic metamaterial absorbers which can be adapted in this frequency window. In our design, each cavity is designed to allow movement of its back plate through a slider, such that we can adjust the volume of the cavity (see Movie S1 in supplementary material). With the slider movement characterized by the parameter $\delta$ (referred to as "slider position"), the geometric parameter $a = a_0 + \delta$, where $a_0$ = 40 mm, and the moving range of $\delta$ is 0-30 mm. Two extreme configurations are considered, the minimum $\delta$ = 0 mm ($a$ = 40 mm) and the maximum $\delta$ = 30 mm ($a$ =70mm). In both configurations, $b$ = 45 mm, $d_{chan}$ = 1.4mm, $d_{slit}$=1.4 mm. The absorption spectra of the two extreme configurations are shown in Fig. 2(b). The experiments (solid line) and simulations (dashed line) achieve a favorable agreement. We also compare their performance with a commercial sound-absorbing foam (Basotect G+, BASF) cut to the same occupied volume of the absorber. It is clear that, around the working frequencies, the absorber has an absorption performance significantly better than the foam. Therefore, it shows that with movement of the sliders, the frequency of the absorption peak is shifted sensitively while maintaining efficient absorption. The efficient absorption is also maintained under air ventilation, as we will discuss below.

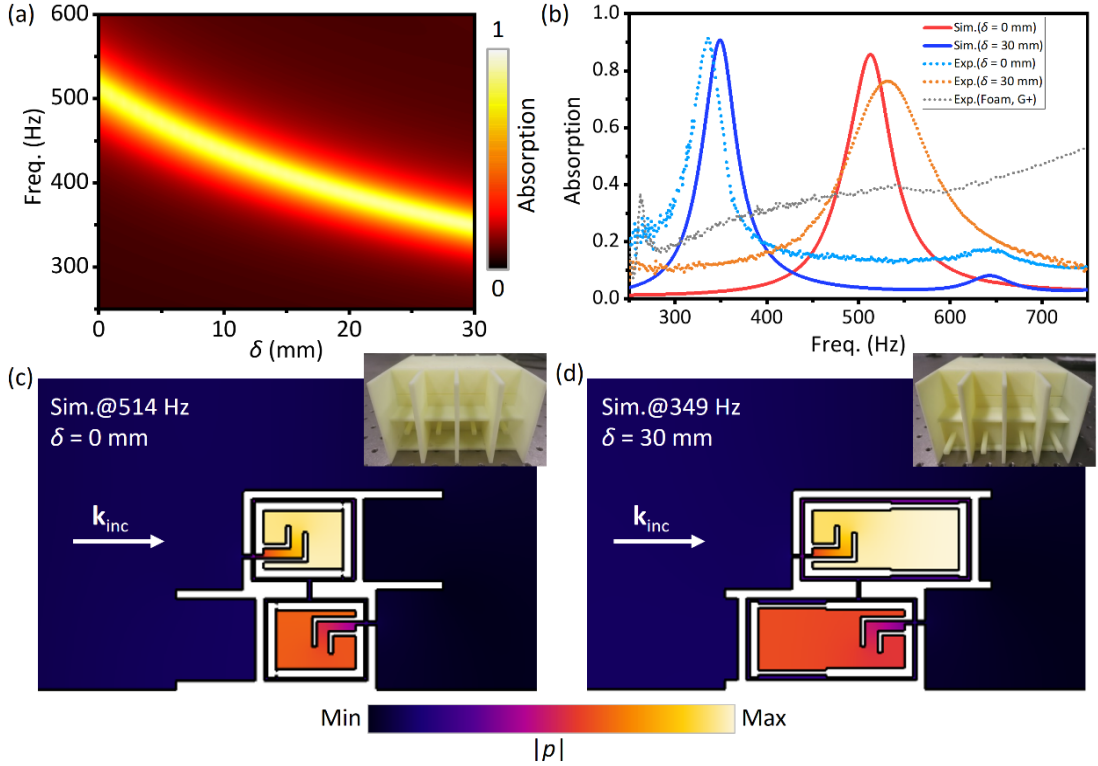

Fig. 2. (a) Simulated spectra of absorption as a function of frequency and geometric parameter *a*. (b) Experimental measurement (dashed lines) and simulation (solid line) absorption spectra of the MTVMA units. For $\delta$ = 0 mm (blue solid line and cyan dashed line), *a* = 80 mm, *p* = 40 mm, $d_{chan}$ = 1.4 mm, $d_{slit}$ = 1.4 mm. For $\delta$ = 30 mm (red solid line and orange dashed line), *a* = 140 mm, *b* = 45 mm, $d_{chan}$ = 1.4 mm, $d_{slit}$ = 1.4 mm. (c), (d) Simulated acoustic pressure field maps on cross section *y* = 0 for $\delta$ = 0 mm and $\delta$ = 30 mm as the frequency is at resonance (514 Hz and 349 Hz), respectively. At resonant frequencies, the acoustic pressures inside cavities for both configurations exhibits strong enhancement. The insets show photographs of the sample configured to $\delta$ = 0 mm and $\delta$ = 30 mm in experiments, respectively.

The sectional views of acoustic pressure fields for the two extreme configurations ($\delta$ = 0 mm and $\delta$ = 30 mm) of the MTVMA are plotted, as shown in Fig. 2(c) and Fig. 2(d), respectively. The pressure amplitude at their resonances (514 Hz and 349 Hz) are shown. In the narrow channels, the sound energy is dissipated and converted into heat under the action of friction, stimulated by the large pressure differences between the inside cavities and outside environment. The upper and lower cavities support identical eigenmodes, and the narrow slits on the cavities make the upper and lower resonant cavities form a weak coupling. Such conditions ensure that the absorption peaks of the two cavities are coalesced rather than split due to radiation couplings[44], so that the total absorption rate at resonances exceeds 50%, as we will discuss below using an effective model.

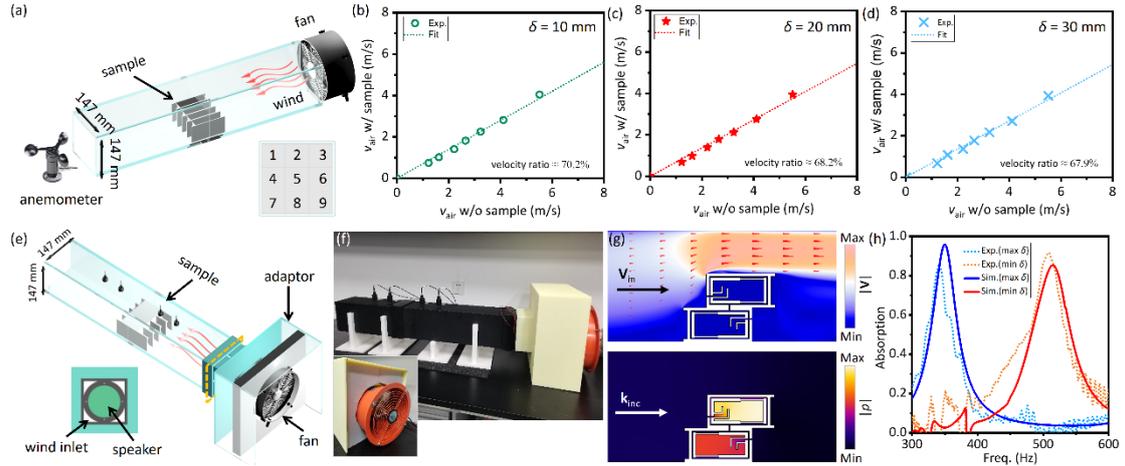

Fig. 3. (a) Schematic of the ventilation characterization system, which measures wind velocities. The inset indicates the nine positions at the tube outlet where the air flow velocities are recorded and averaged. (b)-(d) Experimental measurements (colored scatters) and their linear fittings (colored lines) for air velocities with and without the samples, while the slider is positioned at different positions. The fitted slopes give the wind velocity ratios which characterize the ventilation performance. (e) Schematic of the experiment setup for measuring acoustic simultaneously absorption under air ventilation. Orange dashed rectangle denotes where the speaker is fixed, and its surrounded wind inlet is illustrated in the inset. (f) Photograph for the setup in (e). Inset shows a detailed photograph of the electric fan. (g) Simulated background velocity field ($|\mathbf{V}|$) and acoustic pressure field ($|p|$) on cross section $y = 0$ for $\delta = 30$ mm and the frequency is at resonance (349 Hz). The red arrows denote the background velocity field. (h) Experimental measurement (dashed lines) and simulation (solid line) absorption spectra of the MTVMA units under air ventilation.

The air velocity is ~1.5 m/s.

Further, to investigate ventilation performance of the MTVMA units with movement of the sliders, we characterize their wind velocity ratios, defined as the measured ratios of the air velocities with and without the samples[44,45]. The measurement instruction is shown in Fig. 3(a) and the sample is fixed in the middle of aluminum tube when the air velocity with sample is measured. The electric fan is fixed at the inlet of the aluminum tube and the anemometer is fixed at the outlet. The sample is take out from the tube when the air velocity without sample is measured. We then take averages of the measurements for each configuration, as we measure at nine different positions of the outlet as indicated in the inset in Fig. 3(a). We then perform linear fittings to extract the wind velocity ratios while the slider is positioned at different positions. The results from measurement of air velocities are summarized in Figs. 3(b)-3(d). It can be seen that the linear fittings give efficient wind velocity ratios which are generally around 69%. Therefore, the high-efficiency ventilation of the MTVMAs is validated and also guaranteed during tuning their working frequencies.

Now, there still exists a key issue that whether the MTVMAs can indeed work simultaneously with background air in motion. Previously works usually only investigate the ventilation performance and absorption performance separately. To investigate this issue, we introduce background winds in the impedance through an adaptor, as shown in Fig. 3(e) and Fig 3(f), which illustrate the scheme and actual photograph of the experiment setup. As shown in inset of Fig. 3(e), the wind inlet assures that we perform experimental measurements for acoustic absorption under simultaneous air ventilation. The inlet wind velocity is ~1.5 m/s (See Movie S2 in supplementary material). The simulated background air velocity field and acoustic pressure field at resonance is shown in Fig. 3(g), which reveals that the cavities where acoustic dissipation mainly happens are generally not disturbed by the air flows. Under the condition, the simulated and experimentally measured absorption spectra are shown in Fig. 3(h), which agree quite well. As can be seen, the background wind, at low speed, only introduces noises in the absorption spectra, and the key performances of the MTVMA are firmly retained.

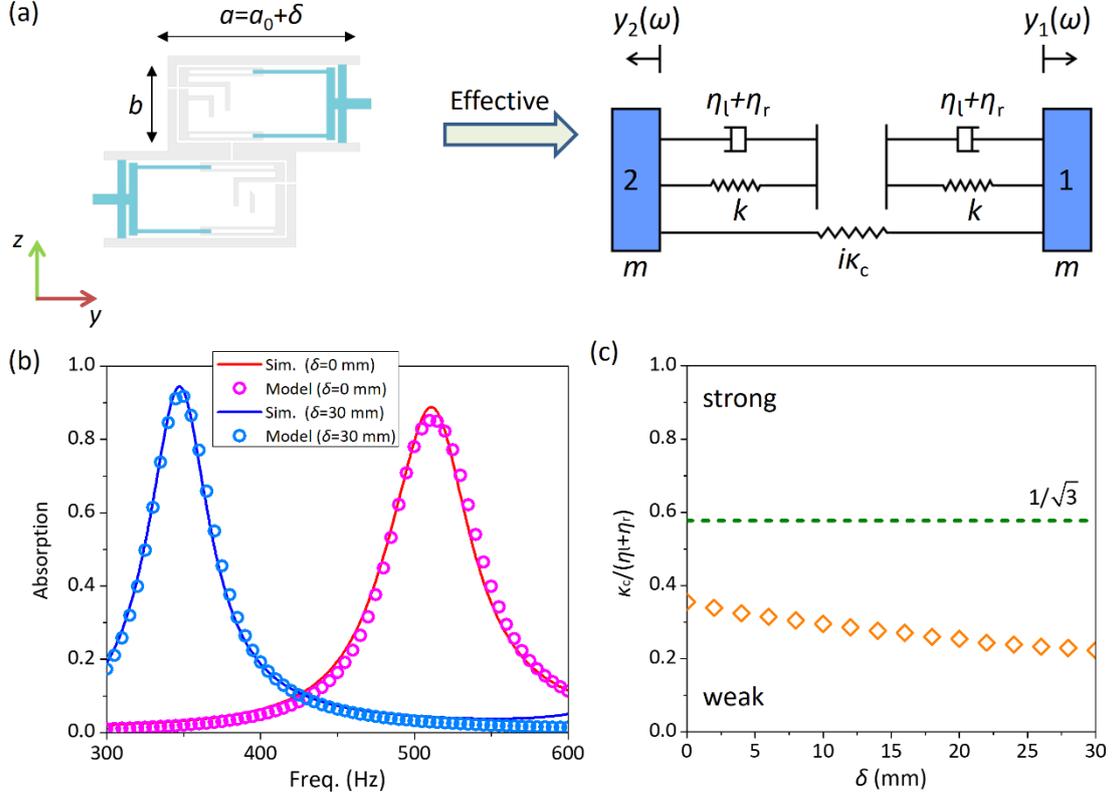

Fig. 4. (a) Schematic of the effective model which describes the acoustic performance of the MTVMA. (b) Absorption spectra for $\delta = 0$ mm and $\delta = 30$ mm from simulations and the effective model, respectively. (c) The ratio $\kappa_c/(\eta_l+\eta_r)$ versus slider position $\delta$, plotted as orange scatters, indicating the strength of the coupling between the resonators versus losses during the movement of the slider. The dashed line indicates the critical value $1/\sqrt{3}$, which distinguishes between the strong ($>1/\sqrt{3}$) and weak ($<1/\sqrt{3}$) couplings. The model parameters $\eta_l$, $\eta_r$ and $\kappa_c$ are retrieved from numerically calculated eigenfrequencies of the MTVMA at different slider positions.

After these experimental validations, now we employ an effective model of coupled lossy oscillators to understand the high-efficiency absorption mechanism of the MTVMAs[44,46]. In fact, an absorption unit can be modeled as a pair of coupled lossy oscillators, as shown in Fig. 4(a), with the effective mass $m$ and the effective stiffness $k$. The radiation loss and the thermal loss are modeled as the damping $\eta_r$ and $\eta_l$, respectively. The dynamic equation of the two oscillators can be written as

$$m\frac{d^2}{dt^2}\begin{bmatrix}y_1\\y_2\end{bmatrix}+\begin{bmatrix}\eta_l+\eta_r & i\kappa_c \\ i\kappa_c & \eta_l+\eta_r\end{bmatrix}\frac{d}{dt}\begin{bmatrix}y_1\\y_2\end{bmatrix}=\begin{bmatrix}F_1\\F_2\end{bmatrix}, \qquad (1)$$

where $y_1$ and $y_2$ are vibration displacements of the oscillators in the effective model, with $F_1$ and $F_2$ denoting the forces on them, and $i\kappa_c$ denotes their radiation coupling. These model parameters can be retrieved from the eigenfrequency simulations, and they obviously depend on the slider position $\delta$. We assume the $e^{-i\omega t}$ time-harmonic convention, and after similar but cumbersome procedures[44], the absorption coefficient in the effective model can be expressed in the frequency domain as (see supplementary material for derivation details)

$$A=\frac{2\eta_l\eta_r^0(1+|Y_{21,0}|^2)}{\eta_0^2[\frac{\omega_0}{\omega}-\frac{\omega}{\omega_0}+2\frac{\mathrm{Im}(\eta_r+i\kappa_c Y_{21,0})}{\eta_0}]^2+4[\eta_l+\mathrm{Re}(\eta_r+i\kappa_c Y_{21,0})]^2}, \qquad (2)$$

with the parameter

$$Y_{21,0}=\frac{2i\kappa_c}{-2(\eta_l+\eta_r)+i\eta_0(\frac{\omega}{\omega_0}-\frac{\omega_0}{\omega})}. \qquad (3)$$

The parameters $\omega_0=\sqrt{k/m}$ is the resonance angular frequency, $\eta_0=2\sqrt{km}$ is the critical damping coefficient, and $\eta_r^0$ is an auxiliary parameter, referred to as the reference radiation loss. To validate the model, we retrieve the model parameters $\omega_0$, $\eta_l$, $\eta_r$ and $\kappa_c$ from eigenfrequency simulations for different $\delta$, and compare the fitted absorption spectra with those from full-wave simulations, as plotted in Fig. 4(b). It is seen that the fitted spectra capture the essence of the numerical results, and hence we can safely describe the ventilated absorber with the effective model. As previously demonstrated[44], in order to achieve an efficient absorption, the model parameters need to satisfy that

$$\kappa_{\mathrm{c}} < \frac{\eta_{\mathrm{l}} + \eta_{\mathrm{r}}}{\sqrt{3}}. \tag{4}$$

Therefore, the ratio $\kappa_{\mathrm{c}}/(\eta_{\mathrm{l}}+\eta_{\mathrm{r}})$ and the critical value $1/\sqrt{3}$ distinguish between the weak coupling ($\kappa_{\mathrm{c}}/(\eta_{\mathrm{l}}+\eta_{\mathrm{r}})<1/\sqrt{3}$, two absorption peaks coalesce) and strong coupling ($\kappa_{\mathrm{c}}/(\eta_{\mathrm{l}}+\eta_{\mathrm{r}})>1/\sqrt{3}$, two absorption peaks split) regimes[44]. To investigate couplings in the MTVMA, we retrieve the model parameters from full-wave simulations, and plot the ratio $\kappa_{\mathrm{c}}/(\eta_{\mathrm{l}}+\eta_{\mathrm{r}})$ in Fig. 4(c). It is clear seen that values of the ratio are significantly smaller than the critical value $1/\sqrt{3}$ within the designed range of slider position $\delta$. Therefore, during the movement of the slider, it is guaranteed that the radiation coupling between the two resonators in an absorption unit is weak enough, such that the absorption peaks from the two resonators coalesce, resulting in an efficient absorption. This is largely attributed to the very small slit width $w_{\mathrm{slit}}$ in our designed metamaterial.

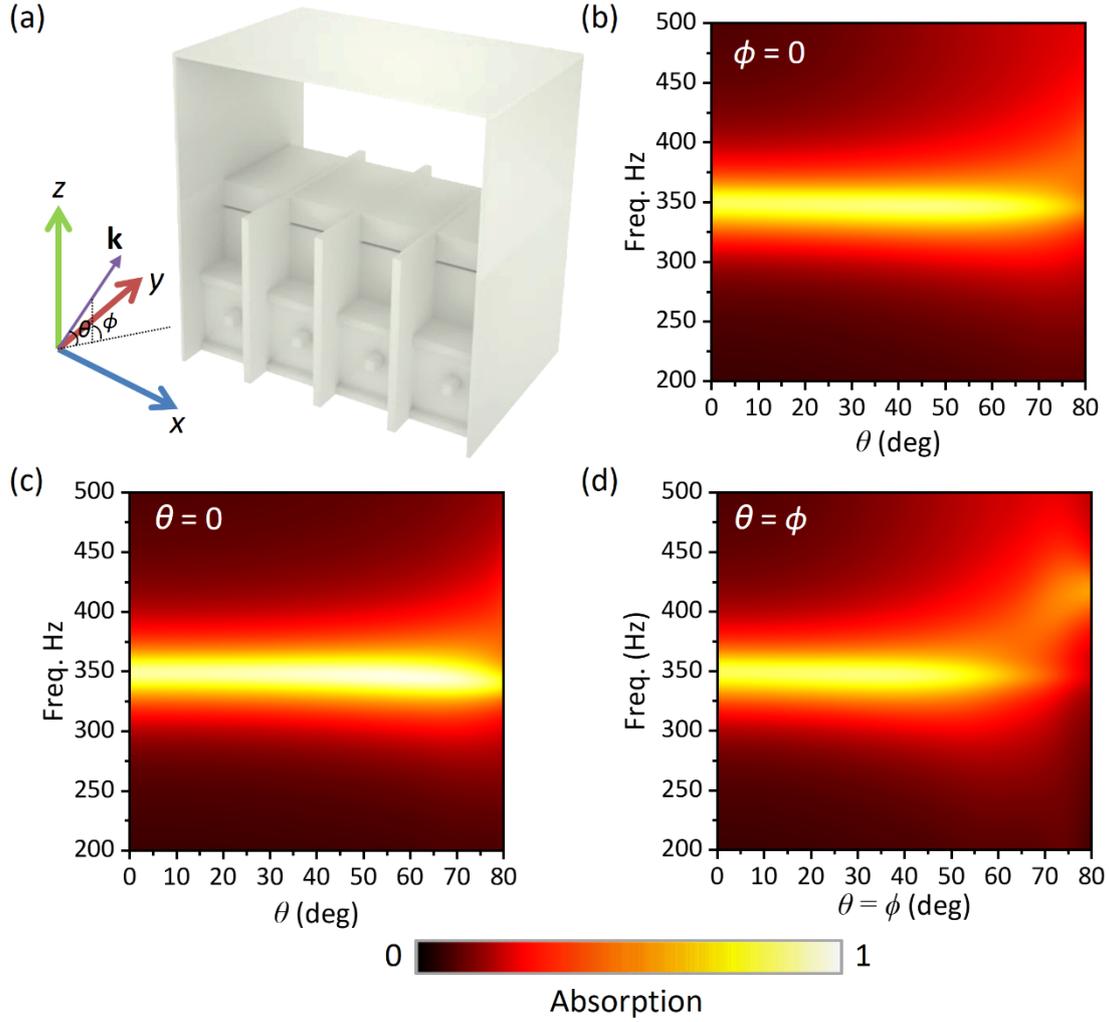

Fig. 5. (a) Supercell in a rectangular lattice that is consisted of four MTVMA units under the oblique incidence and the angles of incidence are defined as $\phi$ and $\theta$ in inset. **k** represents the wave vector. (b)-(d) Simulated spectra of absorption under oblique incidence and the angles are defined as $\theta$ varying, $\phi = 0$ (b), $\phi$ varying, $\theta = 0$ (c), and $\theta = \phi$ varying (d) , respectively.

Finally, we investigate the influence of oblique incidence, and the absorption spectra under different incident angles are calculated from full-wave simulations. The direction of incident sound is characterized by incident angles labeled as $\theta$ and $\varphi$, shown in Fig. 4(a). We simulate the absorption of MTVMA under oblique incidence with varying $\theta$ and $\varphi$, and the results are shown in Figs. 5(b) and 5(c), respectively. The absorption only decreases slightly with the incident angles $\theta$ and $\varphi$ changing from normal incidence to large oblique incidence (> 60°). This is due to the deep subwavelength profile of MTVMA (~ 1/8 wavelength at 350 Hz) which makes the structure relatively insensitive to the incident angle. We also explored the absorption spectra of MTVMA when $\theta = \varphi$. It can be seen that the MTVMA still retains characteristics of the efficient absorption with ventilation under oblique incidence.

To conclude, in this work, we propose and experimentally demonstrate an MTVMA, which is a ventilated metamaterial absorber that can be manually adjusted so as to change the working frequency after being fabricated, meanwhile retaining the desired high performances at low frequencies. The MTVMA is comprised of weakly-coupled Helmholtz resonators, and we can shift working frequency of the MTVMA within the important frequency window of 300-500Hz by simply sliding the sliders. The key to the efficient absorption and ventilation of the MTVMA is the weak coupling between its two Helmholtz resonators, which leads to the coalescence of the absorption peaks of the symmetric and the anti-symmetric modes. Moreover, the tunability of their sound absorption frequencies lies in the sensitive dependence of the resonance frequencies on the cavity geometries that are adjusted by the shift of the sliders. The MTVMA breaks the limit of the previous ventilated sound-absorbing metamaterials that the absorption frequency cannot be changed after being fabricated.

Therefore, the MTVMA has promising application potential in various noise control scenarios, such as duct ventilation and engine systems, since the designers need to handle varieties of noises and ventilation performance is also one of the most concerns. Moreover, the MTVMA paves the way for the development of automatically adjustable sound-absorbing ventilated metamaterials in the future. In addition, the working principle in this work is also suitable for other background fluids.

## Supplementary Material

See Supplementary Material for the animation demonstrating the moving of the sliders on cavities, the demonstration of simultaneous sound absorption under air ventilation, the setup details of simulations in COMSOL Multiphysics, and the derivation of the absorption in the effective model.

## Acknowledgement

This work was supported by the National Key Research and Development Project (2019YFC1906100), Fundamental Research Funds for the Central Universities (2019CDYGYB017), National Natural Science Foundation of China (11974067), Natural Science Foundation Project of CQ CSTC (cstc2019jcyj-msxmX0145, cstc2019jcyj-bshX0042) and Sharing Fund of Chongqing University's Large-scale Equipment, and Hong Kong Research Grants Council (RGC) grant (AoE/P-02/12).

## Data Availability

The data that support the findings of this study are available from the corresponding authors upon reasonable request.